\documentclass[12pt]{article}

\usepackage{epsfig}

\usepackage{epstopdf}
\usepackage[table]{xcolor}

\usepackage[utf8]{inputenc}
\usepackage{graphicx}
\usepackage{amssymb}
\usepackage{amsxtra}
\usepackage{amsmath}
\usepackage[margin=1.0in]{geometry}
\usepackage{booktabs,multirow,tabularx}
\usepackage{slashed}
\usepackage{float}
\usepackage{rotating}
\usepackage{lscape}
\usepackage{color}
\usepackage{hyperref}
\usepackage{enumitem}
\usepackage{ulem}
\usepackage{cite}
\usepackage{subcaption}

\usepackage[Q=yes,pverb-linebreak=no]{examplep}

\makeatletter
\newcommand{\specialcell}[1]{\ifmeasuring@#1\else\omit$\displaystyle#1$\ignorespaces\fi}
\makeatother

\DeclareGraphicsExtensions{.eps}
\graphicspath{{./}}
\newcommand{\pt}{p_{\scriptscriptstyle T}}

\newcommand{\be}{\begin{equation}}
\newcommand{\ee}{\end{equation}}

\newcommand{\mt}{m_t}

\newcommand{\gsim}{\gtrsim}

\newcommand\sss{\scriptscriptstyle}

\newcommand{\mh}{m_{ \sss H}}
\newcommand{\mz}{m_{ \sss Z}}

\def\beq{\begin{equation}}
\def\bea{\begin{eqnarray}}
\def\eeq{\end{equation}}
\def\eea{\end{eqnarray}}
\def\beqnl{\begin{align}}
\def\endal{\end{align}}

\makeatletter
\newcommand{\normalorbold}{%
  \ifnum\pdf@strcmp{\math@version}{bold}=\z@ bx\else m\fi
}
\makeatother

\begin{document}\color{black}
\begin{titlepage}
\nopagebreak

\renewcommand{\thefootnote}{\fnsymbol{footnote}}
\vspace{1cm}
\begin{center}
  {\Large \bf \color{magenta} Gluon Fusion Production at NLO: Merging
    the Transverse Momentum and the High-Energy  Expansions}
  
\bigskip\color{black}\vspace{0.6cm}
 {\large\bf Luigi Bellafronte$^a$\footnote{email: lui.bellafronte@usc.es},
       Giuseppe Degrassi$^b$\footnote{email: giuseppe.degrassi@uniroma3.it},
       Pier Paolo Giardino$^{a}$\footnote{email: pierpaolo.giardino@usc.es},
       Ramona Gr\"{o}ber$^{c}$\footnote{email: ramona.groeber@pd.infn.it},
       Marco Vitti$^b$\footnote{email: marco.vitti@uniroma3.it} }
     \\[7mm]     
       {\it (a) Instituto Galego de F\'isica de Altas Enerx\'ias,
         Universidade de Santiago de Compostela, 15782 Santiago de Compostela,
 Galicia-Spain}\\[1mm]
{\it (b) Dipartimento di Matematica e Fisica, Universit{\`a} di Roma Tre and \\
 INFN, sezione di Roma Tre, I-00146 Rome, Italy}\\[1mm]
{\it (c) Dipartimento di Fisica e Astronomia 'G.~Galilei',
  Universit\`a di Padova and INFN, sezione di Padova, I-35131 Padova, Italy}\\     
\end{center}

\bigskip
\bigskip
\bigskip
\vspace{0.cm}

\begin{abstract}
The virtual corrections to $gg\to HH$ and $gg\to ZH$ are analytically evaluated
combining an expansion in the small transverse momentum of the final particles
with an expansion valid at high energies. The two expansion
methods
describe complementary regions of the phase space and we 
merge their results, extending the range of validity of both expansions
using Padé approximants. We  show that this approach can
reproduce the available numerical results retaining the exact top quark mass
dependence with an accuracy well below the 1\% level.  Our results allow a
fast and flexible evaluation of the virtual corrections of the considered
processes. Furthermore, they are available in different renormalisation
schemes of the top quark mass.
\end{abstract}
\vfill  
\end{titlepage}    

\setcounter{footnote}{0}

\section{Introduction}
\label{sec1}
At the Large Hadron Collider (LHC) gluon fusion  is the most relevant
production mechanism for Higgs physics in  single-Higgs production, $gg \to H$,
and  pair production, $gg \to HH$, and plays an important role in the
production of a Higgs boson in
association with a $Z$ boson, $gg \to ZH$.  Precise predictions
for such processes are necessary in order to measure the properties of
the Higgs boson accurately.
Since the Higgs boson is a colorless particle, these processes are all
loop-induced and mediated by a heavy particle, mainly the top quark.
As known since more than twenty years, QCD corrections to the Born result are
very large and the perturbative expansion converges slowly
(see for example \cite{Catani:2003zt,Ahrens:2008qu}). This implies
that the knowledge of the higher-order QCD corrections is very important.
However, the evaluation  of these corrections is extremely challenging already
at next-to-leading order (NLO), since it requires  the computation of
two-loop diagrams. 

In this paper we are going to consider the virtual corrections to 
Higgs production via gluon fusion at the NLO level. In general, the degree of
difficulty in the evaluation of loop diagrams grows
with the number of energy scales present in the diagram. 
In the case of single-Higgs production
the relevant diagrams feature a triangular topology and, consequently, depend
upon only two scales, namely the Higgs mass, $\mh$, and the top
mass\footnote{ All the quarks but the top are assumed to be
  massless.}, $\mt$.  In this case, the functional dependence of the result
upon the top mass   can be expressed in terms of one single
variable, $\mh^2/\mt^2$. Due to this ``simplified'' one-scale situation,
exact analytic results for the NLO corrections are available since many years
\cite{Spira:1995rr,Harlander:2005rq,Aglietti:2006tp,Anastasiou:2006hc}.

In the case of processes with two particles in the final state the situation
is more complicated. 
Indeed these processes receive contributions not only from triangle diagrams, that can
be calculated adapting the exact analytic results obtained for single-Higgs
production, but also from box-topology diagrams.
In pair production, $gg \to HH$, the box diagrams depend
upon four  scales, namely $\hat{s},\, \hat{t},\, \mt, \mh$,
where $\hat{s},\, \hat{t}$, and $\hat{u}$ are the
Mandelstam variables which satisfy the condition
\be
\hat{s} + \hat{t} + \hat{u} = 2 \,\mh^2 ~.
\ee
Concerning associated production,  $gg\to ZH$, a fifth energy scale is present,
i.e.~the mass of the $Z$ vector boson, $m_Z$. 

Exact analytic results for two-loop box diagrams with several energy scales
cannot be derived  with the
present computational technology. Instead, usually two different strategies are
followed in order to  evaluate  the two-loop box contribution in Higgs
production via gluon fusion.
a) A fully  numerical exact evaluation
\cite{Borowka:2015mxa,Borowka:2016ehy, Baglio:2018lrj, Baglio:2020ini,
  Chen:2020gaew}.
b) An approximate analytic evaluation that takes advantage of  hierarchies
among the various energy scales present in the diagrams, in order to
reduce the number of scales in the problem. Thus, its  validity is restricted to
specific regions of the phase space. The method used is based on the
expansion of the diagrams in terms of ratios of small energy scales vs.~large
energy scales, in order to obtain  a result that retains an exact dependence
upon the large energy scales.  Concerning the small ones,
in order to simplify further the evaluation,  expansions in term of ratios
between small energy scales is often used.

The former strategy, although accurate, is very demanding from a computational
point of view, requiring  a high degree of optimization in order to obtain a
result in a ``reasonable'', although usually quite long, computer time.
Furthermore, this approach is not very flexible with respect to the
modification of the input parameters. 

Strategy b) provides  accurate results valid in specific regions of
the phase space without requiring heavy computational work, i.e.~in a
short computer time. Examples of this approach of
evaluating the two-loop box contribution are:
\begin{enumerate}
\item[i)] The infinite-top-mass limit \cite{Dawson:1998py, Altenkamp:2012sx}
refined by the inclusion of powers in the large  top-mass expansion (LME)
\cite{Grigo:2013rya,Grigo:2015dia,Degrassi:2016vss,Hasselhuhn:2016rqt}.
Here, $\mt$ is assumed to be the large energy scale while
$\hat{s}, \: \hat{t},\: \mh$, and in associated production also $\mz$,
are considered to be the small ones. Thus, the validity of
this approach is restricted to phase-space regions where
$\hat{s}/(4 \mt^2) \leq 1$.\\
\item[ii)] The evaluation via an expansion in the transverse momentun, $\pt$,
  of the final-state particles \cite{Bonciani:2018omm,Alasfar:2021ppe}.
Here, $\hat{s}$ and $\mt$ are assumed to be the large energy scale while 
$ \mh, \:  \mz$ and $\pt$, that can be traded for $\hat{t}$,  are considered
to be the small ones.
The validity of this approach is restricted to phase-space regions where
$ | \hat{t} |/(4 \mt^2) \lesssim 1$. \\
\item[iii)] The evaluation via a high-energy (HE) expansion
\cite{Davies:2018ood,Davies:2018qvx,Davies:2020drs}.
Here $\hat{s},\: \hat{t}$ are assumed to be the large energy scale while 
$\mt,\:\mh$ and $\mz$,  with $\mt \gg \mh,\, \mz$, are considered to be the
small ones.
The validity of this approach is restricted to phase-space regions where
$ | \hat{t} |/(4 \mt^2) \gtrsim 1$.\\
\item[iv)] The evaluation via an expansion in terms of small external masses
  \cite{Xu:2018eos,Wang:2020nnr,Wang:2021rxu}. Here $\hat{s}, \: \hat{t},\:\mt$
  are assumed to be the large energy scale
while  $\mh$ and $\mz$   are considered to be the small ones. This approach
basically covers the entire phase space of the considered processes. However, since the reduction of scales in this
approach is minimal, one ends up with the evaluation of Master Integrals (MIs) 
that are much more complicated than those  appearing in the
i)--iii) cases. As a consequence the evaluation of the box contribution in any
point of the phase space requires a longer computer time than in the approaches
i)--iii).
\end{enumerate}
As an alternative approach, refs. \cite{Grober:2017uho,Davies:2019roy}
proposed to  reconstruct the full result from its
LME version, supplemented by the non-analytic part of the diagrams near the
top threshold, via a conformal mapping and Pad\'e approximants.

In this paper we propose an alternative way to derive the full
top-mass dependence in Higgs production via gluon fusion, based on the
merging of the $\pt$ expansion in ii) with the HE expansion in iii)
that individually are valid in complementary regions of the phase
space.  
Since the numerical evaluations of the two expansions are quite fast from a
computational point of view, our proposal allows a fast evaluation of
the virtual corrections to Higgs production via gluon fusion that is
accurate in the entire phase space.

The key point of our analysis is to extend the
fixed-order results both in the $\pt$ expansion
\cite{Bonciani:2018omm,Alasfar:2021ppe} and in the HE expansion
\cite{Davies:2018qvx,Davies:2020drs} up to or beyond
their border of validity, i.e.~$\hat{t} \simeq 4 \mt^2$, in order to merge the two analytic approximations. This is done by constructing a [1/1] Pad\'e approximant for the
$\pt$-result and a [6/6] Pad\'e approximant for the HE-result. We point out that the extension of the HE expansion via Pad\'e approximants has been already considered in refs.~\cite{Davies:2018qvx,Davies:2020drs}.

The paper is organized as follows. In section \ref{sec2} we introduce
the different expansions as well as the method for combining them.
In section \ref{sec3} we validate the method at LO both at the level
of form factors and at the level of the partonic cross
section, with a focus on the $gg \to HH$ process. In section \ref{sec4} we present the merging of the
two expansions at NLO.  In order to show the flexibility of our approach with
respect to the modification of the input parameters, in the same section we present the
result for the two-loop virtual contribution in $gg \to HH$ and $gg \to ZH$
for two different choices of the top quark mass, namely the on--shell and the
$\overline{\text{MS}}$ mass. Finally, we conclude in section \ref{sec5}.

\section{Method}
\label{sec2}
We start by considering the process
$g (p_1)\, g(p_2) \rightarrow 3(p_3)\, 4 (p_4)$, where 3 and
4 are two neutral\footnote{Dealing with neutral particles in the final state implies the absence of mixed top-bottom diagrams, as they would appear e.g.~in $gg\to W^+W^-$ production, for which the proposed method cannot be straightforwardly applied.} particles with
masses $m_3$ and $m_4$, respectively.  Taking all momenta to be
incoming,  the partonic Mandelstam variables are
\begin{equation}
\hat{s} = (p_1 + p_2)^2, \qquad \hat{t} = (p_1+p_3)^2, \qquad \hat{u} = (p_2+p_3)^2,
\end{equation}
and the transverse momentum $\pt$ of the final-state particles can be written
as
\begin{equation}
\pt^2 = \frac{\hat{t} \hat{u} - m_3^2 m_4^2}{\hat{s}}.
\end{equation}
As suggested in refs.~\cite{Bonciani:2018omm,Alasfar:2021ppe}, if the
amplitude of the process
is written in terms of (anti)symmetric form factors with respect to the exchange $\hat{t}
\leftrightarrow \hat{u}$, then it is sufficient to discuss only the
forward contribution to the cross section. Therefore, in the following we
will always assume that $|\hat{t}| \leq |\hat{u}|$ and that
\begin{equation}
  \hat{t} = -\frac{1}{2} \left( \hat{s}- m_3^2 - m_4^2 -
  \sqrt{\lambda(\hat{s},m_3^2,m_4^2)-4 \hat{s}\, \pt^2} \right),
\end{equation}
where $\lambda(a,b,c) = a^2 + b^2 +c^2 -2 a b -2a c-2b c$ is the K\"all\'en
function.

In the forward regime, the validity of both the $\pt$ and HE expansions is
limited by the condition
\begin{equation}
|\hat{t}| \simeq 4 \mt^2,
\label{eq:tcondition}
\end{equation}
i.e.~for any fixed value of $\hat{s}$, the $\pt$ expansion provides
reliable results when $|\hat{t}| \lesssim 4 \mt^2$ while the HE
expansion is accurate for $|\hat{t}| \gtrsim 4 \mt^2$, if the fixed $\hat{s}> 4 \mt^2$. 
However, we find that in the vicinity of the point $|\hat{t}|=4 \mt^2$
the fixed-order results in the $\pt$ expansion and in the HE
expansion are both divergent (see fig.~\ref{fig:LO}). As a
consequence, a straightforward combination of the $\pt$-expanded and
the HE-expanded results cannot allow for an accurate description of
the above region, and this fact prevents a full coverage of the phase
space. We point out that this situation does not change substantially
when higher orders in both the expansions are computed.

Alternatively, the convergence of the expanded results can be improved
by considering the respective Pad\'e approximants. Indeed, starting from
a given Taylor expansion of an exact function $f(x)$ around $x=0$ up
to the first $r$ terms

\begin{equation}
f(x) \simeq \sum_{k=0}^{r-1} c_k x^k ,
\label{eq:taylorexp}
\end{equation}
it is possible to construct the associated Pad\'e approximant, defined as
\begin{equation}
[m/n](x) = \frac{p_0 + p_1 x + \dots + p_m x^m}{1+ q_1 x + \dots  q_n x^n},
\label{eq:mnpade}
\end{equation}
provided that $m+n+1=r$. Specifically, by Taylor-expanding the
r.h.s.~of eq.~\eqref{eq:mnpade}, the $\{p_i,q_j\}$ coefficients of the
Pad\'e approximant can be written in terms of the $c_k$ ones known from
eq.~\eqref{eq:taylorexp}, by solving a system of linear equations.  Usually,
$[m/n]$ Pad\'e approximants such that $m=n$ give the best
improvement in the convergence of the original Taylor expansion, and
we consider only these combinations in our study. In  the
$\pt$-expanded  results, at NLO,  only the first three terms in
eq.~\eqref{eq:taylorexp} are known and therefore we are limited to
construct  a $[1/1]$ Pad\'e approximant (we will refer
to this as the $\pt$-Pad\'e).  Instead the availability of many terms in
the HE-expansion results allows to consider several $[n/n]$
approximants (defined as HE-Pad\'e).

When calculating the $\pt$-Pad\'e, care is to be taken in the treatment
of the expansion parameters. As discussed in
refs.\cite{Bonciani:2018omm,Alasfar:2021ppe}, not only the $\pt$ but
also the masses of the external particles are understood as small
parameters. Since these are all treated on the same footing with
respect to the large scales set by $\hat{s}$ and $\mt$, we can write
the general expression for a $\pt$-expanded form factor $F$
in the amplitude
in terms of a scaling parameter $x$
\begin{equation}
  F(x) = \sum_{N=0}^2 x^N  \sum_{i+j+k=N}  c_{ijk} ~ (\pt^2)^i (m^2)^j (\Delta_m)^k
  \equiv \sum_{N=0}^2 x^N c_N
\label{eq:ffscalingx}
\end{equation}
where $m$ is interpreted as $m_H$ and $m_Z$ for $gg \rightarrow HH$
and $gg \rightarrow ZH$, respectively, and $\Delta_m =
(m_H^2-m_Z^2)/2$ is included only for the $ZH$ case (see
ref.~\cite{Alasfar:2021ppe}).  Starting from eq.~\eqref{eq:ffscalingx}
we can then obtain the corresponding $[1/1]$ Pad\'e approximant with
respect to the limit $x\rightarrow 0$
\begin{equation}
[1/1](x) = \frac{p_0 + p_1 x}{1 + q_1 x},
\label{eq:11pade}
\end{equation}
with
\[
p_0 = c_0 \qquad p_1 = c_1 - \frac{c_0 c_2}{c_1} \qquad q_1 =  -\frac{c_2}{c_1},
\]
 and subsequently set $x=1$ in eq.~\eqref{eq:11pade}.

We want to clarify a possible source of ambiguity concerning the limit
of validity of the $p_T$ expansion. Indeed, while in the previous
works we suggested that this expansion is valid for $p_T^2 \lesssim
4m_t^2$, as the comparison at LO between the $\pt$-expanded
and exact result seems to indicate,
  in this paper we follow a more conservative approach and we
  consider as  limit of validity for the $\pt$ expansion
  $|\hat{t}| \lesssim 4m_t^2$. Additionally, we checked that the same
complementarity for the $\pt$ and HE  expansions can be observed when choosing
$p_T^2 = 4m_t^2$ as limit of validity.

We now discuss the procedure adopted to construct the Pad\'e
approximants from the HE expansion. Following the prescription of
ref.~\cite{Davies:2019dfy} (see also
\cite{Davies:2020lpf,Wellmann:2020rxl}), we initially arrange the
various orders $F^{(i)}$ of the HE expansion for a given form factor
as follows
\begin{equation}
  F(x) = F^{(0)} + \sum_{l=1}^{L}\left( F^{(2l-1)} m_t^{(2l-1)} + F^{(2l)} m_t^{(2l)}
  \right) x^l = \sum_{l=0}^L d_l x^l,
\label{eq:heseries}
\end{equation} 
where orders related to odd powers of $m_t$ are grouped with the
orders related to the next even power.  Then, we construct $[n/n]$
approximants in $x$ with $2n=L$ from eq.~(\ref{eq:heseries}) 
using the analytic expressions available in
\cite{kithhpage,kitzhpage}, and setting $x=1$.  We remark that
our Pad\'e approximants are obtained in a fully symbolic way, whereas in
refs.~\cite{Davies:2019dfy,Davies:2020lpf} all the kinematical
quantities are fixed to the respective numerical values before the
Pad\'es are constructed in $x$. Furthermore, in comparison to
refs.\cite{Davies:2019dfy,Davies:2020drs}, we only studied $[n/n]$
HE-Pad\'es up to $n=6$. In those references Pad\'es with $n>6$ were also
considered in order to extrapolate the results in the region
$|\hat{t}| < 4 \mt^2$, for a fixed $\hat{s}$, and characterize the
relative uncertainties of different $[m/n]$ Pad\'es. In our case,
because the region $|\hat{t}| < 4 \mt^2$, is more accurately described
by the results of the $\pt$ expansion, we find that a [6/6] HE-Pad\'e
is more than enough to perform the merging with the $\pt$-result and,
at the same time, to describe accurately the high-energy region.

The $\pt$-Pad\'e and the HE-Pad\'e extend the range of
  validity of each expansion beyond its limit. As
discussed in the next section, the $\pt$ and the HE Pad\'es are
accurate enough to bridge the gap around the phase-space region
$|\hat{t}|\simeq 4\mt^2$. Then, an accurate approximation of the exact
result for any phase-space point ($\hat{s},\hat{t}$) can be obtained
by choosing as switching point between the Pad\'e-improved
expansions any point in the region $|\hat{t}| \sim 4 \mt^2$. For simplicity
we choose to use the $\pt$-Pad\'e when $|\hat{t}| < 4m_t^2$ and the HE-Pad\'e when
$|\hat{t}| \geq 4m_t^2$, for any fixed value of $\hat{s}$. We recall that in our discussion we just consider the forward region $|\hat{t}| \leq |\hat{u}|$, while  the result in the complementary phase-space region
is obtained   using the symmetry of our form factors  under $\hat{t} \leftrightarrow \hat{u}$. Noticing that, when $|\hat{t}| \leq |\hat{u}|$, 
the maximum absolute value of $\hat{t}$ as a function of $\hat{s}$ is given by
$|\hat{t}|_\text{max} =1/2 (\hat{s} - m_3^2 - m_4^2)$ our choice corresponds to
using the $\pt$-Pad\'e up to the partonic energy 
$\hat{s}_c = 8 \mt^2+m_3^2+m_4^2$. In this energy region
($\sqrt{\hat{s}_c} \simeq 500$ GeV for $gg \rightarrow HH$ and
$gg \rightarrow ZH$) at the LHC more than 2/3 of the hadronic cross section is
concentrated.

\section{The $\pt$ and HE expansions vs the  exact results at LO}
\label{sec3}
In this section we assess the reliability of our merging procedure by
studying how well the combination of the $\pt$-Pad\'e and the HE-Pad\'e
can reproduce the exact LO results for $ HH$ and $ZH$ production via
gluon fusion.  For the sake of simplicity, we discuss in detail only
the $gg \rightarrow HH$ process, but we verified that similar conclusions can be
drawn for $gg \rightarrow ZH$.

Using the same notation of ref.~\cite{Degrassi:2016vss}, we recall that the
amplitude for $gg \rightarrow HH$ can be expressed as
\begin{equation}
  A^{\mu \nu} = \frac{G_\mu}{\sqrt{2}} \frac{\alpha_S(\mu_R)}{2 \pi} \delta_{a b}
  T_F \hat{s} [A_1^{\mu \nu} F_1 + A_2^{\mu \nu} F_2], \label{eq:Amunu}
\end{equation}
where $F_1$ and $F_2$ are the form factors associated to the spin-0
and spin-2 projectors, respectively. Both triangle and box
diagrams contribute to $F_1$
\begin{equation}
F_1 = F_\triangle \frac{3 \mh^2}{\hat{s}-\mh^2} + F_\square,
\end{equation} 
whereas the $F_2$ form factor receives contribution only from boxes.
Our goal is to improve the evaluation of the box
contributions,
therefore we focus on the
discussion of $F_\square$ and $F_2$.

\begin{figure}
\begin{subfigure}{.5\textwidth}
  \centering
\hspace*{-0.8cm}  \includegraphics[width=1.2\linewidth]{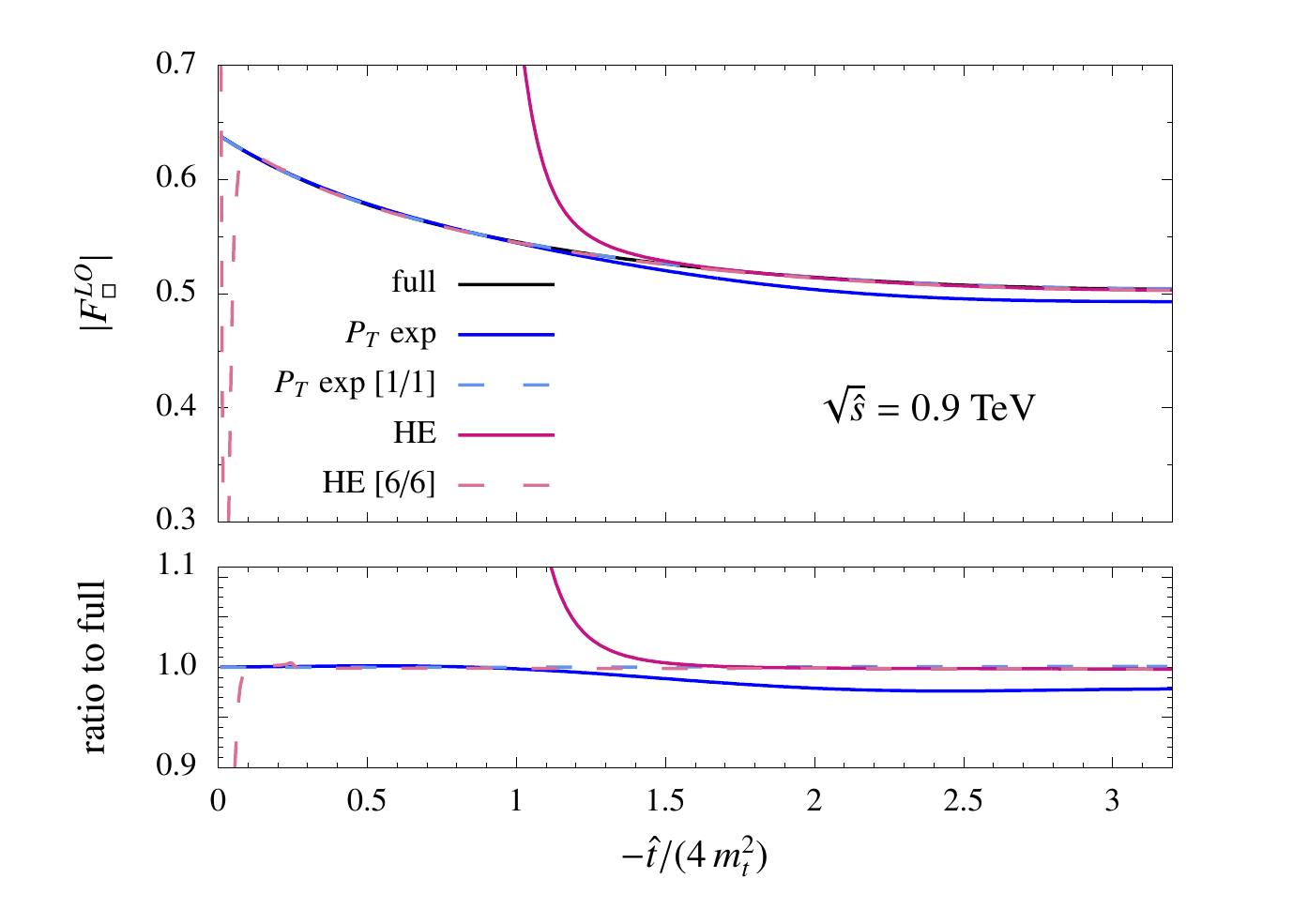}
  \caption{}
  \label{fig:lof900}
\end{subfigure}%
\begin{subfigure}{.5\textwidth}
  \centering
  \includegraphics[width=1.2\linewidth]{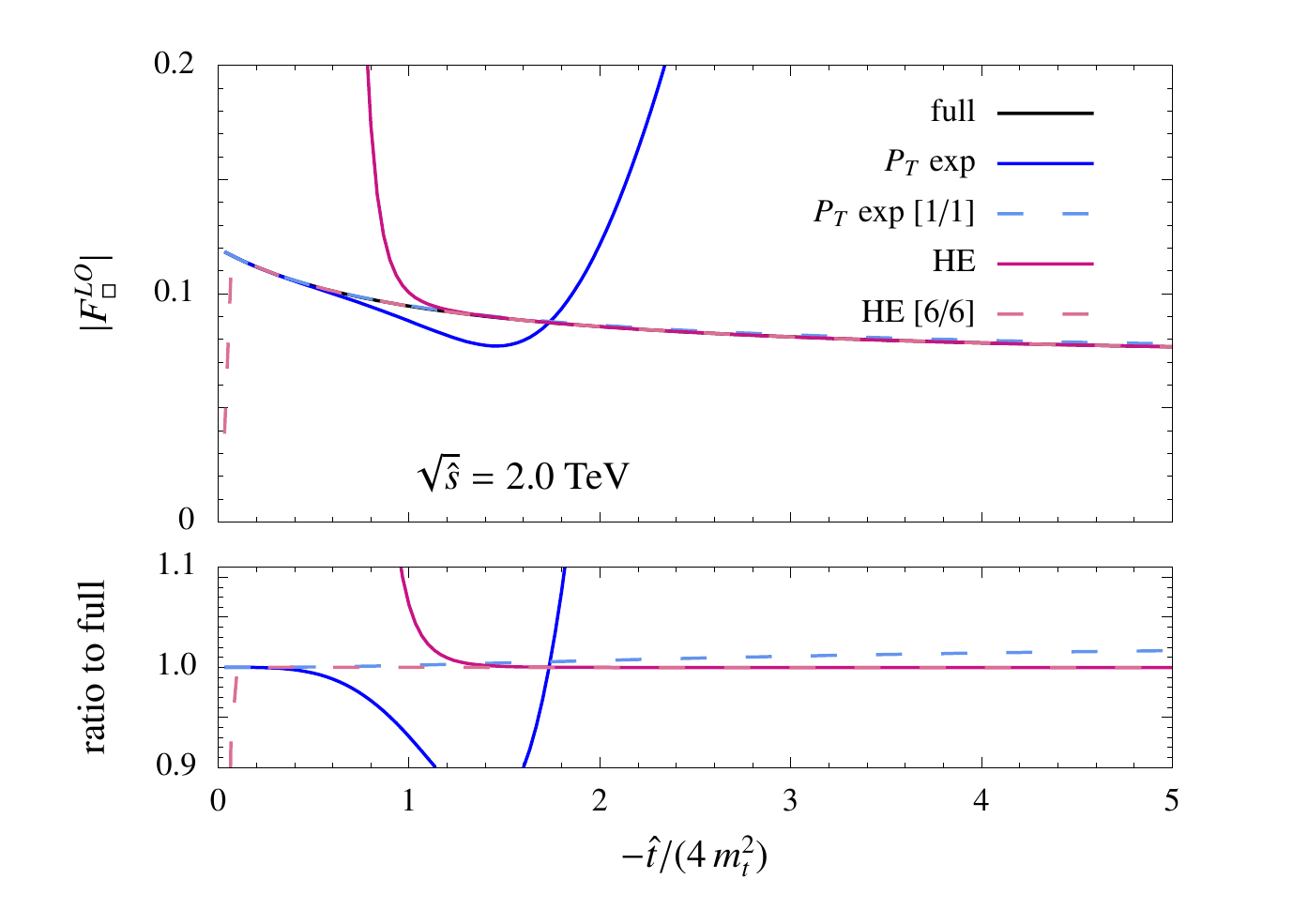}
  \caption{}
  \label{fig:lof2000}
\end{subfigure} \\
\begin{subfigure}{.5\textwidth}
  \centering
\hspace*{-0.8cm}   \includegraphics[width=1.2\linewidth]{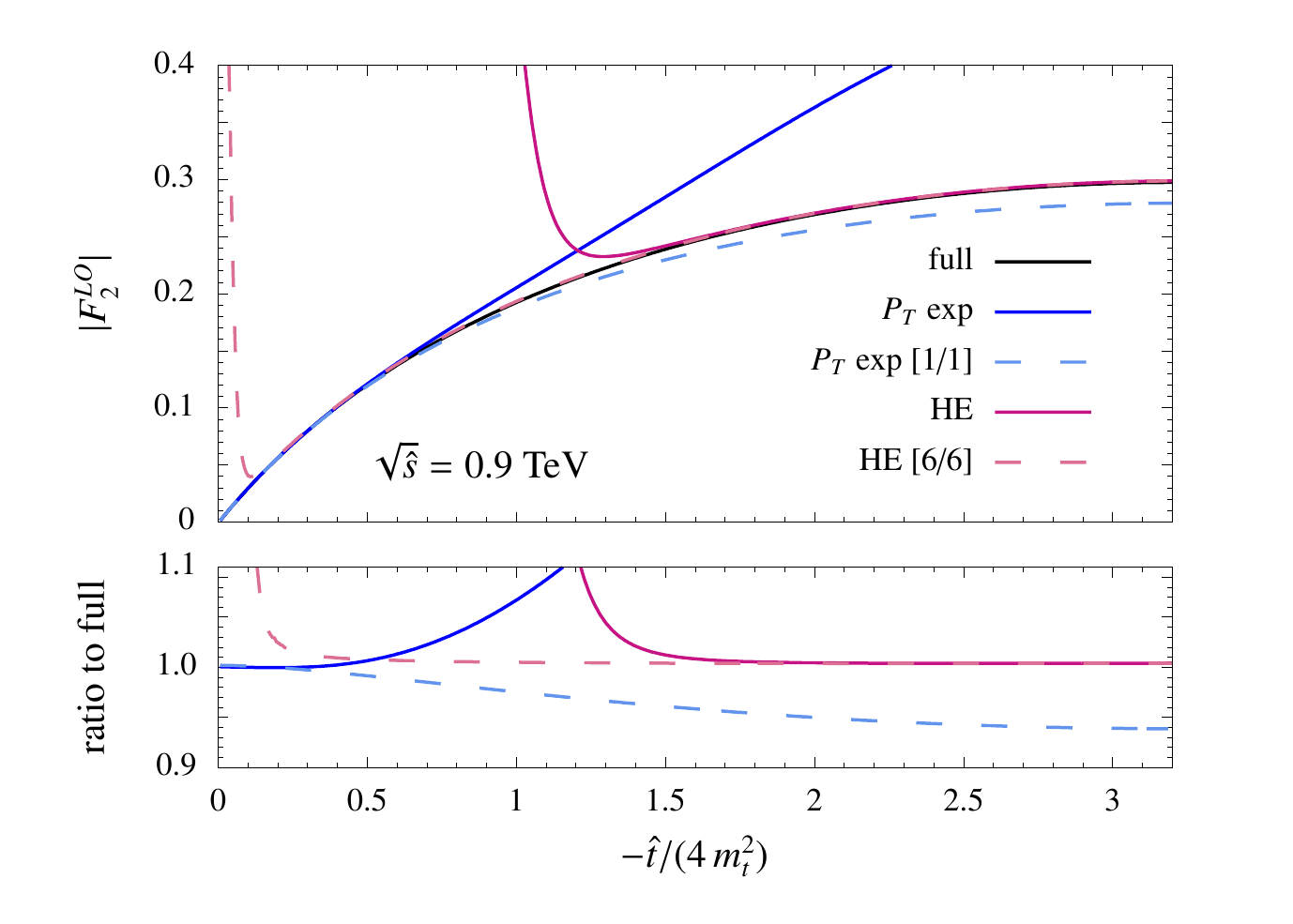}
  \caption{}
  \label{fig:log900}
\end{subfigure}
\begin{subfigure}{.5\textwidth}
  \centering
  \includegraphics[width=1.2\linewidth]{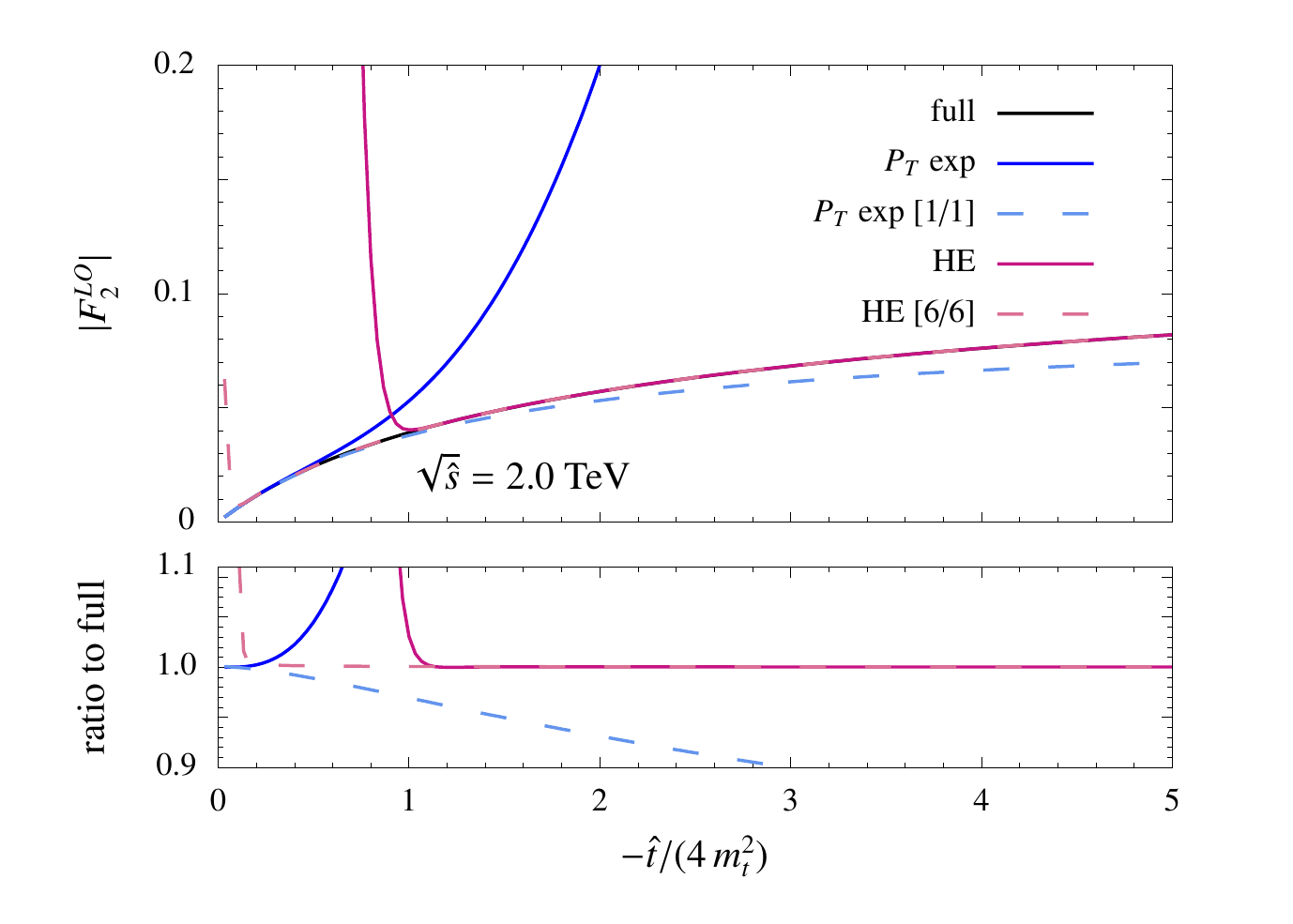}
  \caption{}
  \label{fig:log2000}
\end{subfigure}
\caption{Modulus of the box form factors contributing to $gg
  \rightarrow HH$ at LO, for a fixed value of (a,c) $\sqrt{\hat{s}}
  =0.9$ TeV and (b,d) $\sqrt{\hat{s}} =2$ TeV. In the upper part of
  each plot, the exact prediction (solid black line) is shown together
  with the $\pt$ and HE expansions (solid blue and purple lines,
  respectively) and with the [1/1] $\pt$ and [6/6] HE Pad\'e
  approximants (dashed light blue and pink lines, respectively). The
  bottom part of each plot shows the ratio of the above results to the exact
  prediction.}
\label{fig:LO}
\end{figure}

The LO results for these form factors, denoted as $F_\square^{LO}$ and
$F_2^{LO}$, are shown
in fig.~\ref{fig:LO}, for fixed values of the partonic center-of-mass
energy.
Only large values of $\hat{s}$ are shown in fig.~\ref{fig:LO}
because for small $\hat{s}$ values the $\pt$-expanded results
are very accurate \cite{Bonciani:2018omm}. The
$\pt$-expanded and HE-expanded results are represented by the blue and
purple solid lines, respectively, and they deviate from the exact
result, shown as a solid black line, at $|\hat{t}|/4m_t^2 \simeq 1$,
as anticipated in the previous section. The light blue dashed line
stands for the [1/1] $\pt$-Pad\'e, while the pink dashed line represents
the [6/6] HE-Pad\'e. One can see that the Pad\'e results show an improved
convergence with respect to the fixed-order expansions. 
The bottom part of the plots in fig.~\ref{fig:LO} shows  the ratio of the
expanded and Pad\'e results to the exact one. Indeed, fig.~\ref{fig:LO}(a,b)
shows that in the case of $F_\square^{LO}$ for
$|\hat{t}|/4m_t^2 = 1$ the differences of the Pad\'e results with
respect to the exact prediction are negligible.
For the $F_2$ form factor, whose contribution to the cross-section is
much smaller than the one of the $F_1$ form factor, the difference is
always below 5\%,  see fig.~\ref{fig:LO}(c,d).  We notice that,
when comparing the accuracies of the Pad\'e approximants, larger
discrepancies can be attributed to the $\pt$-Pad\'e. 
Indeed, being the latter a [1/1] Pad\'e, it is expected to be a less refined approximation than the [6/6] HE-Pad\'e. Still, in the case of $F_\square^{LO}$ the differences between the two Pad\'e near $|\hat{t}|/4 m_t^2 = 1$ are negligible.
We also notice that, as
$\hat{s}$ increases, larger values of $|\hat{t}|$ are allowed by the
kinematics, and the relative importance of the HE expansion increases.

\begin{figure}[t]
\begin{subfigure}{.5\textwidth}
  \centering
\hspace*{-0.8cm}  \includegraphics[width=1.2\linewidth]{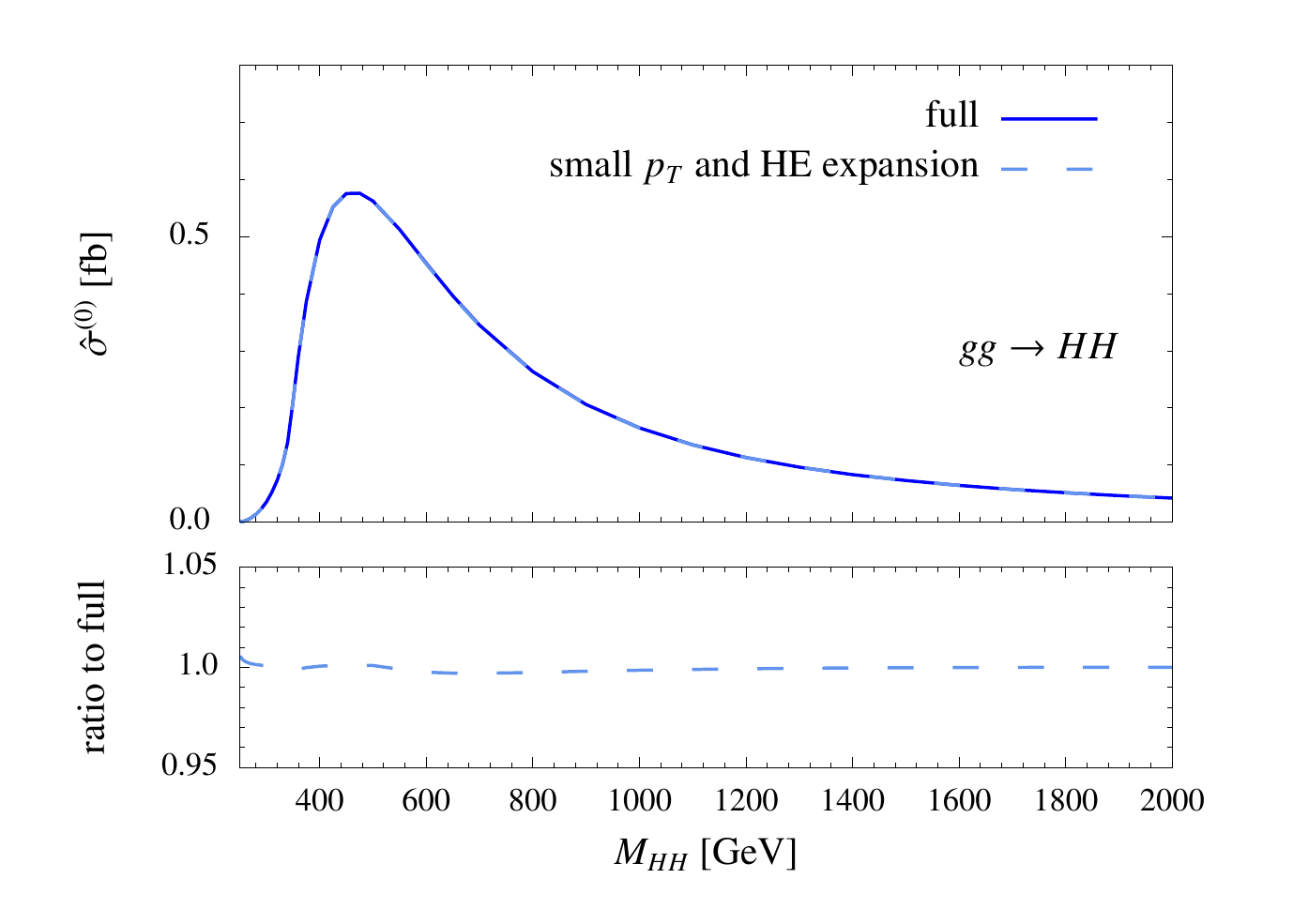}
  \caption{}
  \label{fig:lof900}
\end{subfigure}%
\begin{subfigure}{.5\textwidth}
  \centering
  \includegraphics[width=1.2\linewidth]{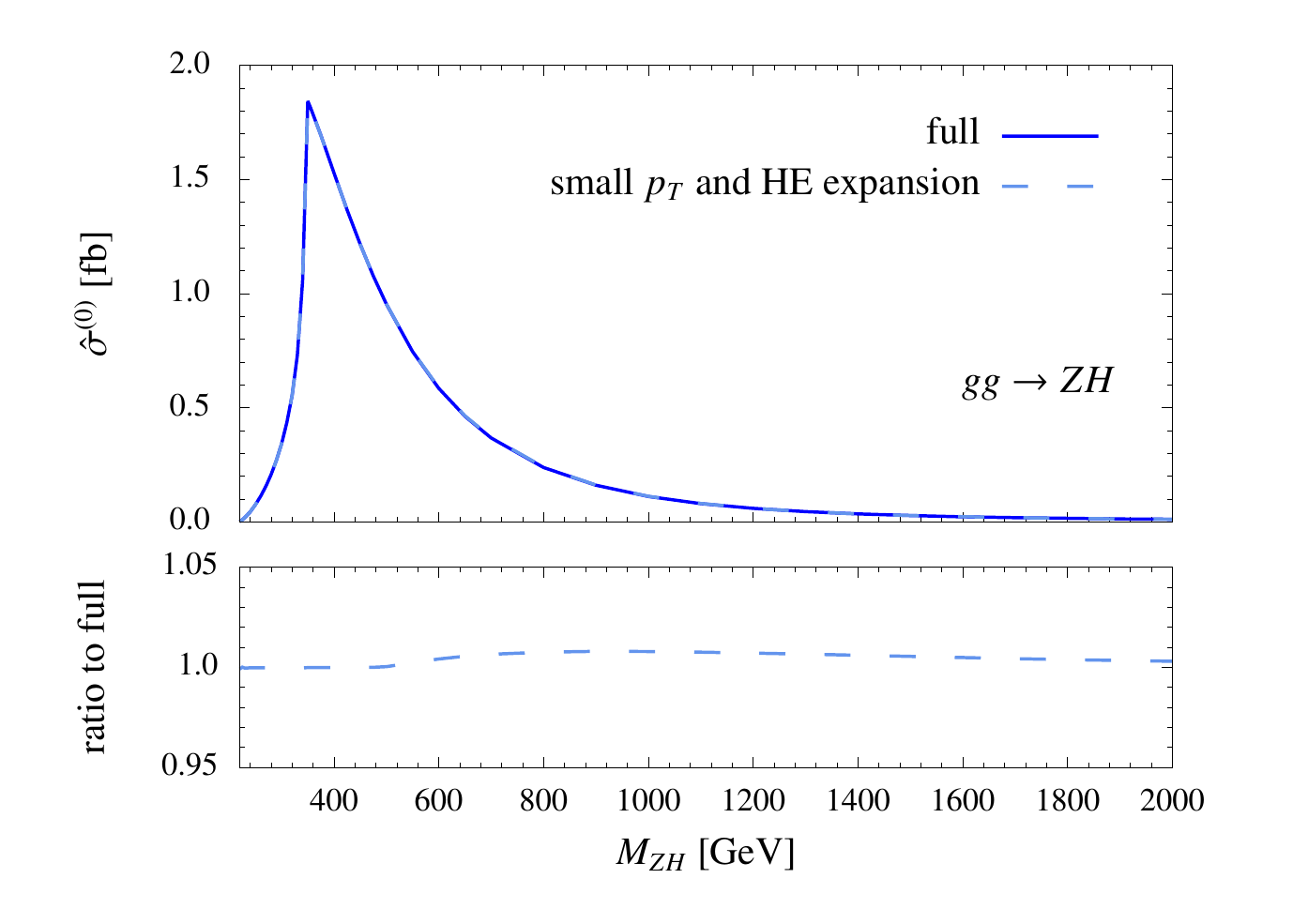}
  \caption{}
  \label{fig:lof2000}
\end{subfigure}
\caption{Partonic cross section at LO for (a) $gg\rightarrow HH$ and
  (b) $gg\rightarrow ZH$. The upper part of each plot shows the exact
  prediction (solid line) together with the merging of the $\pt$ and
  HE Pad\'e approximants (dashed line). The bottom part of each plot shows the ratio of the
  merged result to the exact prediction.}
\label{fig:xsecLO}
\end{figure}

The improvement in convergence provided by the Pad\'e approximants is
such that the merging of the two results discussed in the previous section can reproduce the exact
prediction with good accuracy for every value of $\hat{t}$,
for any $\hat{s}$.  While we refrain from showing more examples here, 
we note that we studied the behaviour of all the box
contributions to $gg\rightarrow HH$ and $gg \rightarrow ZH$ at several
values of $\hat{s}$. We explicitly checked that, among the various
possibilites, a [6/6] HE-Pad\'e is more than enough for an accurate
merging. Furthermore, we observed that the value $|\hat{t}|=4 m_t^2$
is a good choice as a merging point for the $\pt$ and HE Pad\'e
approximants.  The high level of accuracy of our merging method can be observed in fig.~\ref{fig:xsecLO},
where the partonic cross section at LO is shown for $gg \rightarrow
HH$ and $gg \rightarrow ZH$. One can see that deviations of the
combination of the $\pt$- and HE-Pad\'e with respect to the exact
prediction never exceed 1\%.

\section{Merging the $\pt$ and HE expansions at NLO}
\label{sec4}
In the previous section we showed that the merging of the $\pt$- and
HE-Pad\'e can accurately reproduce the exact LO prediciton. In this
section we present the merging of the NLO $\pt$-expanded and
HE-expanded results improved by the respective Pad\'e approximants.
\begin{figure}
\begin{subfigure}{.5\textwidth}
  \centering
\hspace*{-0.8cm}  \includegraphics[width=1.2\linewidth]{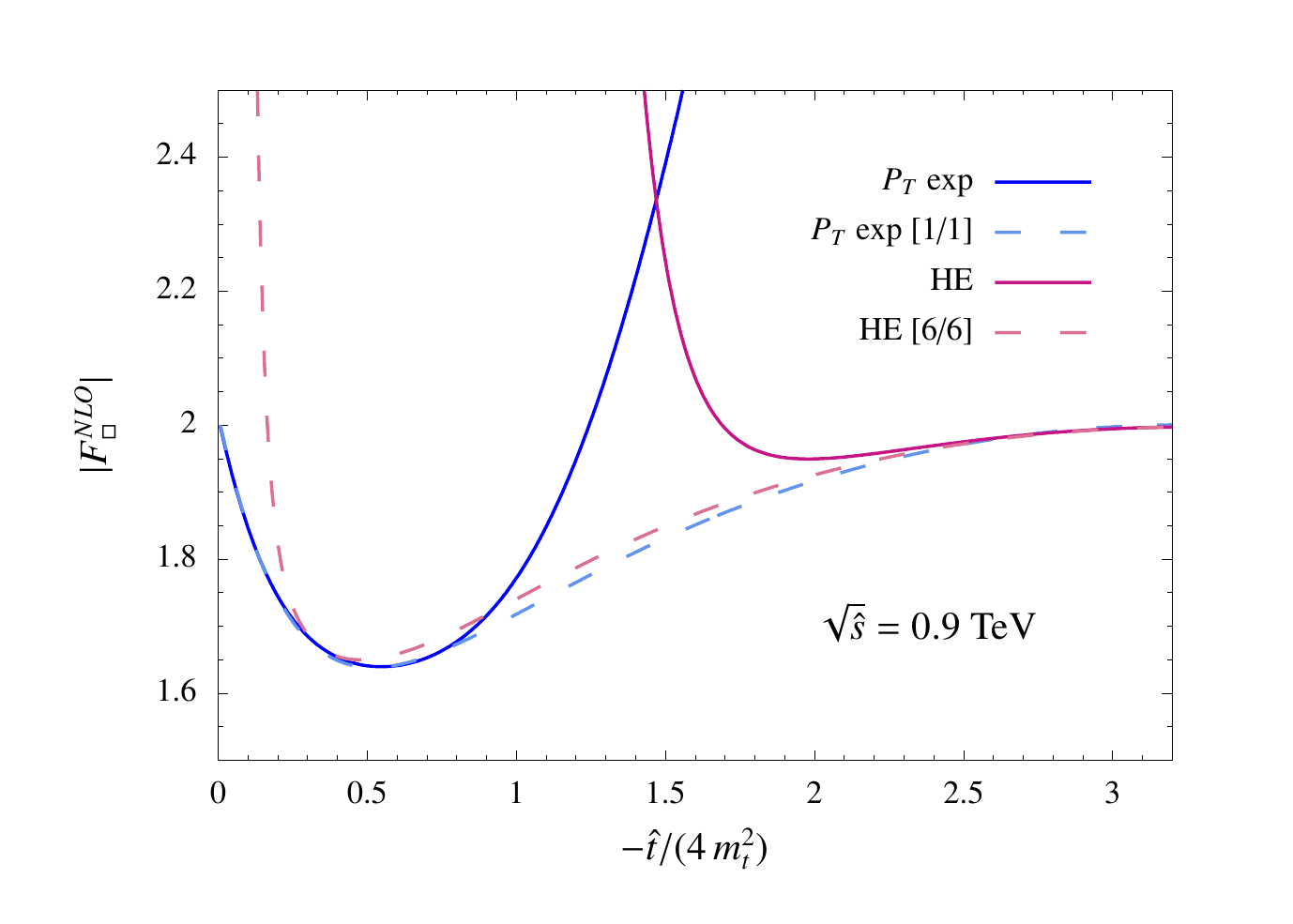}
  \caption{}
  \label{fig:nlof900}
\end{subfigure}%
\begin{subfigure}{.5\textwidth}
  \centering
  \includegraphics[width=1.2\linewidth]{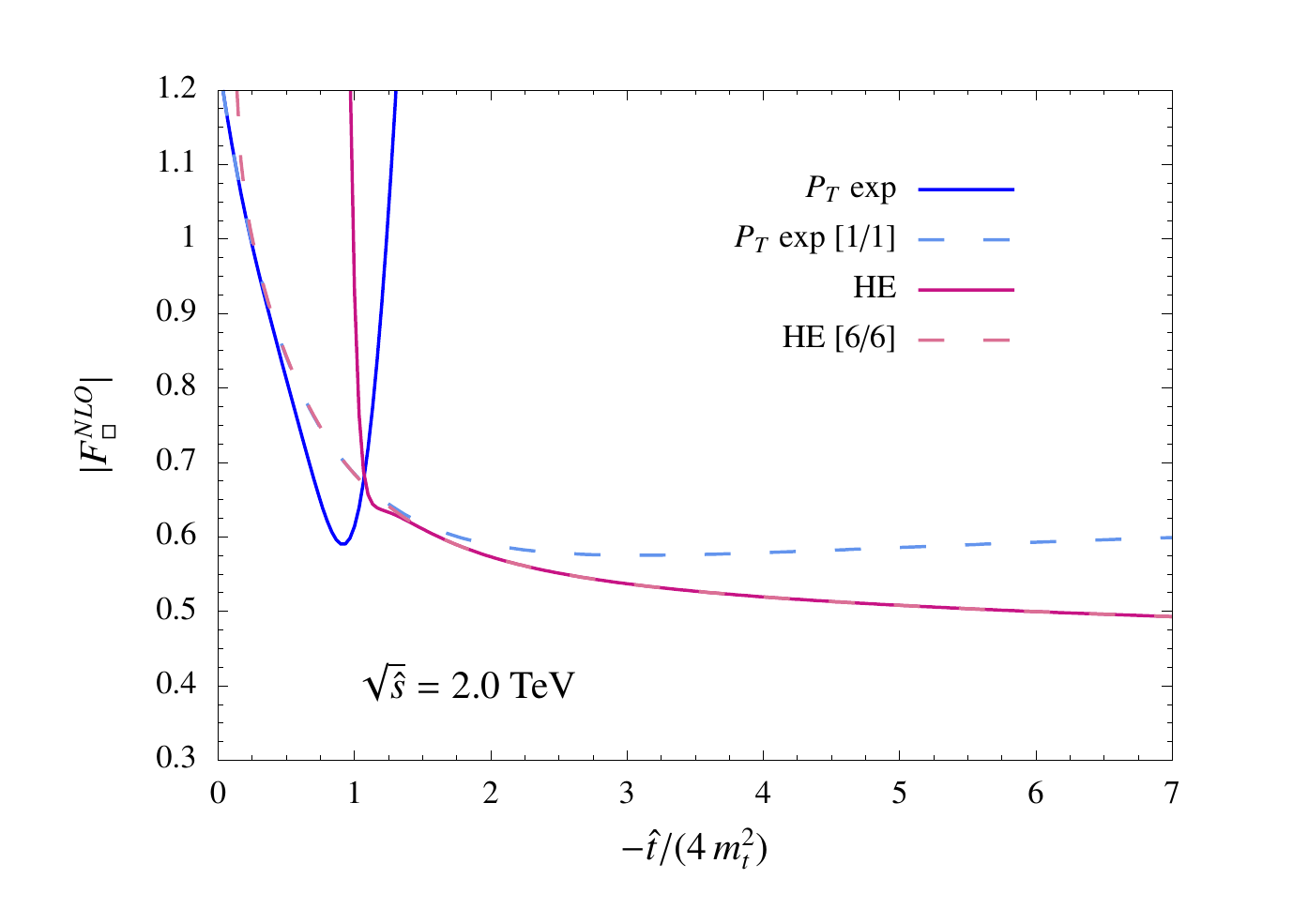}
  \caption{}
  \label{fig:nlof2000}
\end{subfigure} \\
\begin{subfigure}{.5\textwidth}
  \centering
\hspace*{-0.8cm}  \includegraphics[width=1.2\linewidth]{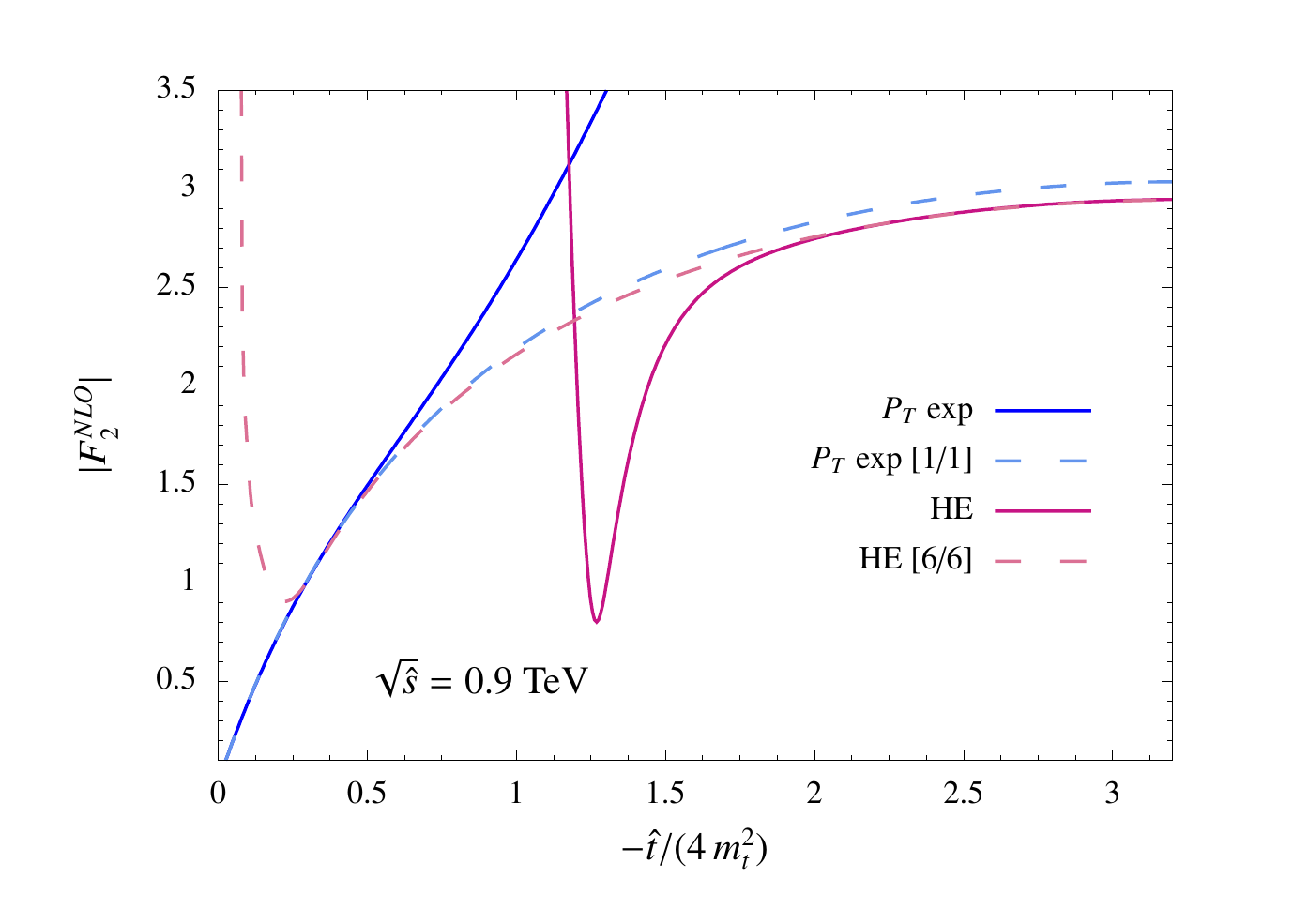}
  \caption{}
  \label{fig:nlog900}
\end{subfigure}
\begin{subfigure}{.5\textwidth}
  \centering
  \includegraphics[width=1.2\linewidth]{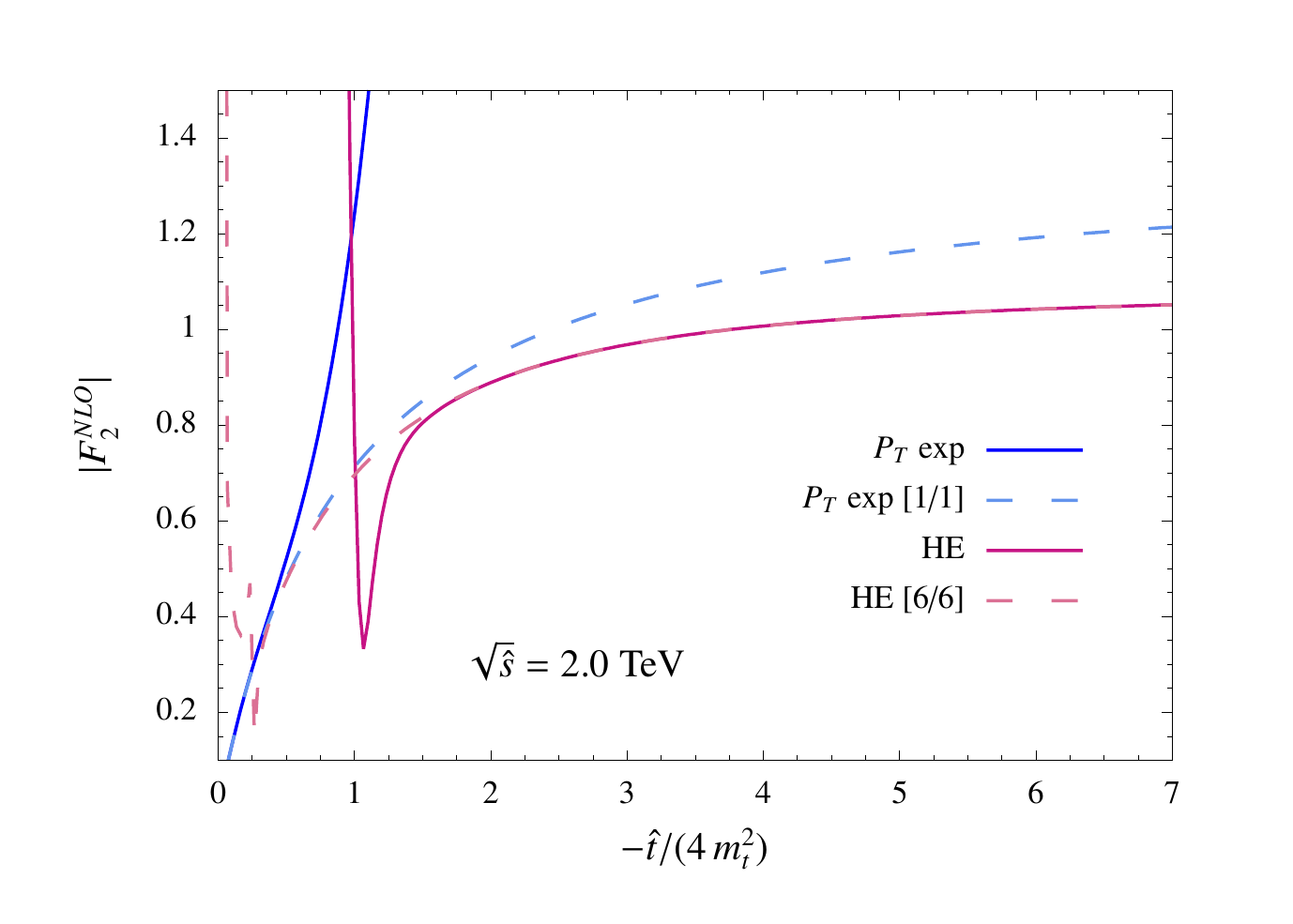}
  \caption{}
  \label{fig:nlog2000}
\end{subfigure}
\caption{Modulus of the box form factors contributing to $gg
  \rightarrow HH$ at NLO, for a fixed value of (a,c) $\sqrt{\hat{s}}
  =0.9$ TeV and (b,d) $\sqrt{\hat{s}} =2$ TeV.  The $\pt$ and HE
  expansions are shown as solid blue and purple lines, respectively,
  while the [1,1] $\pt$- and [6,6] HE-Pad\'e are shown as
  dashed light blue and pink lines, respectively.}
\label{fig:NLO}
\end{figure}

In fig.~\ref{fig:NLO} the NLO contributions to $F_\square$ and $F_2$
are shown\footnote{$F_\square^{NLO}$ and $F_2^{NLO}$ are the form
factors as defined in eq.~\eqref{eq:Amunu} but do not contain the
double triangle diagrams that can be expressed in terms of products of
one-loop integrals and as such are computed analytically in exact
top-mass dependence \cite{Degrassi:2016vss}. }.  The relative
behaviour of the various approximations is analogous to what we
observed at LO.  For low values of $|\hat{t}|$ the fixed-order and
Pad\'e-improved $\pt$-expanded results agree well. Increasing the
value of $|\hat{t}|$ up to the merging region, $|\hat{t}| \sim 4
\mt^2$, the $\pt$-Pad\'e becomes close to the Pad\'e-improved HE
expansion. For values above $|\hat{t}|=4\mt^2$ the $\pt$ and
HE Pad\'e approximants show small deviations as expected.  The NLO
study shows the same qualitative behaviour as the LO one.  This makes
us confident that the proposed merging procedure works well also at
NLO.

\begin{figure}
\begin{center}
\includegraphics[width=0.7\linewidth]{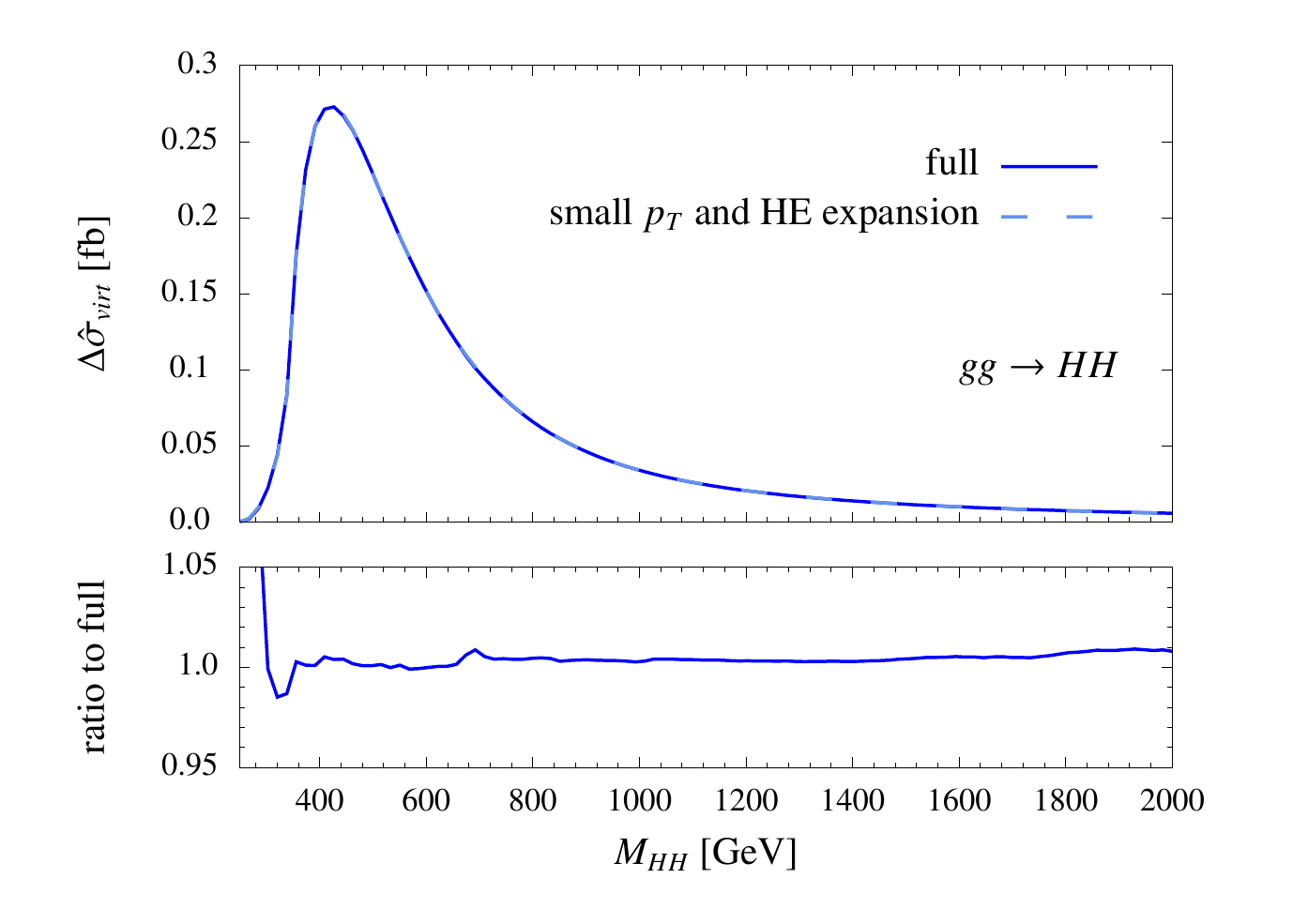}
\caption{Upper pannel: $\Delta \hat{\sigma}_{virt}$ using the 
numerical grid provided in ref.~\cite{Davies:2019dfy} (dark blue line) 
and our combination of HE expansion and small $\pt$ expansion 
(light blue dashed line). The lower panel shows the ratio of the two lines. \label{fig:HHNLO}}
\end{center}
\end{figure}

We now compare  our evaluation of the virtual
corrections for the di-Higgs production process\footnote{We
note that for $ZH$ production no public code including the results of
the full computation \cite{Chen:2020gaew} is currently
available. Hence we refrain from making any comparisons for $ZH$
production.} with the numerical result provided as a grid in
  ref.~\cite{GitHub}. This reference summarizes the work of ref.~\cite{Davies:2019dfy}, where
  the numerical calculation in exact
  top-mass dependence
of ref.~\cite{Borowka:2016ehy}  was  supplemented by the result in the
HE expansion of ref.~\cite{Davies:2018qvx}.
The comparison is done on the quantity
\begin{equation}
  \Delta \hat{\sigma}_{virt}=\int_{\hat{t}_-}^{\hat{t}_+}
  \frac{\alpha_s}{32 \pi^2} \frac{1}{\hat{s}^2}\mathcal{V}_{fin}d\hat{t},
\end{equation}
where the finite part of the virtual corrections $\mathcal{V}_{fin}$
is defined as in ref.~\cite{Grober:2017uho}.  The results are shown in
fig.~\ref{fig:HHNLO}. The grid of ref.~\cite{GitHub}  shows very good agreement
with our results at every
invariant mass, except for the first few bins at low $M_{HH}$. The reason is a
large uncertainty of the numerical grid on the low $M_{HH}$ bins,
that are described by only a few points in the numerical grid due to their small contribution to the total cross section.  For moderate and large $M_{HH}$ we observe
differences below 1\%.

Finally, we show that our merging approach
  is flexible with respect to the modification of the input parameters
  by computing the virtual corrections for various renormalisation
  schemes of the top quark mass.
It was noted in refs.~\cite{Baglio:2018lrj, Baglio:2020ini} that the di-Higgs
production process suffers from a large uncertainty associated to the
renormalisation scheme of the top quark mass.
In particular, an uncertainty on the NLO cross section between $+4\%$
and $-18\%$ \cite{Baglio:2020wgt} is related to the change from the
on-shell renormalisation scheme to the $\overline{\text{MS}}$ scheme
for the top mass, with the latter evaluated at different values of the
renormalisation scale ($m_t$, $M_{HH}$ and $M_{HH}/4$).
The results presented so far have been calculated using the on-shell  scheme
for the top  mass, however the form factors in the
$\overline{\text{MS}}$ scheme can be obtained by simply shifting our result
according to:
\begin{equation}
F_i^{NLO, \overline{\text{MS}}}=F_i^{NLO,\text{OS}}- 
\frac{1}{4}\frac{\partial F_i^{LO}}{\partial m_t^2} \Delta_{m_t^2}
\label{eq:MSbarshift}
\end{equation}
with $i=\Delta,\Box,2$ and 
\begin{equation}
  \Delta_{m_t^2}= 2m_t^2 C_F \left[-4+3 \log\left( \frac{m_t^2}{\mu^2}\right)
                 \right]\,.
\end{equation}
Notice that the numerical values for the top quark mass have to be
adjusted to their $\overline{\text{MS}}$ values, which we evaluate
following refs.~\cite{Melnikov:2000qh,Carena:1999py}.  Since the LO results are
available analytically, we can first build the Pad\'e approximants of
the $\pt$- or HE-expanded $F_i^{NLO,\text{OS}}$ form factors and
then calculate the shift to the the $\overline{\text{MS}}$ scheme
using the full LO form factors in eq.~(\ref{eq:MSbarshift}). Alternatively,
we can use the expanded LO form factors to perform the shift in
eq.~(\ref{eq:MSbarshift}) order by order in the $\pt$-expanded and the
HE-expanded results, and
then build Pad\'e approximants on the $F_i^{NLO,
  \overline{\text{MS}}}$ form factors. The difference between these
two approaches turns out to be well below the 0.5 permille level
everywhere except near the top-mass threshold where the difference
is at the percent level.  For the $gg\to ZH$ process, the shift to the
$\overline{\text{MS}}$ scheme can be applied in analogy to
eq.~\eqref{eq:MSbarshift} on the associated form factors\footnote{The
$\Delta \hat{\sigma}_{virt}$ for $gg \rightarrow ZH$ was defined as in
ref.~\cite{Alasfar:2021ppe}.}.  For the triangle 
contributions we always use the results available in full top-mass
dependence.

\begin{figure}[t]
\begin{subfigure}{.5\textwidth}
  \centering
\hspace*{-0.8cm}  \includegraphics[width=1.2\linewidth]{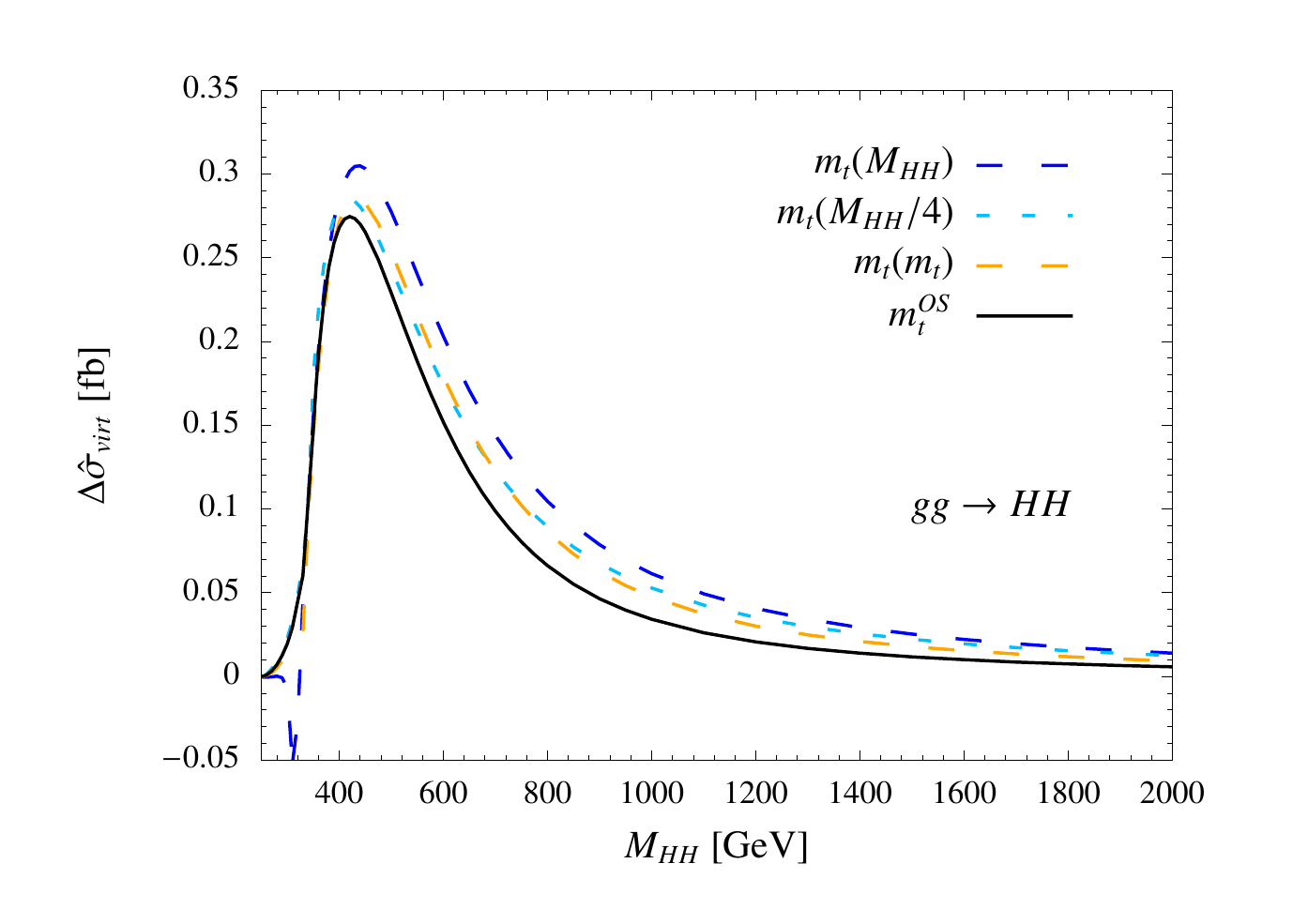}
  \caption{}
  \label{fig:lof900}
\end{subfigure}%
\begin{subfigure}{.5\textwidth}
  \centering
  \includegraphics[width=1.2\linewidth]{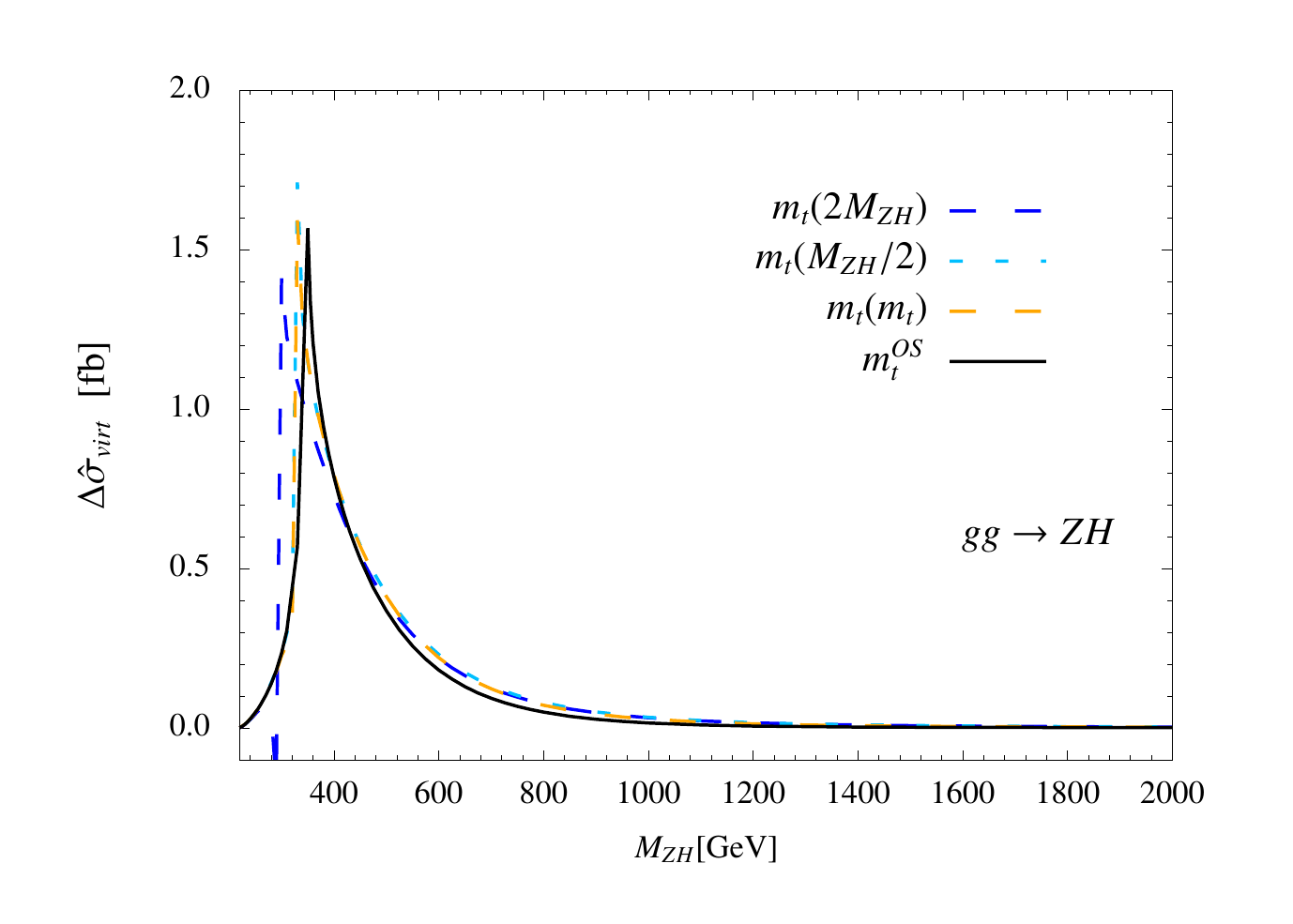}
  \caption{}
  \label{fig:lof2000}
\end{subfigure}
\caption{The integrated virtual corrections in different renormalisation 
schemes for the top quark mass for (a) $gg\rightarrow HH$ and
  (b) $gg\rightarrow ZH$. 
  The dashed lines show $\Delta \hat{\sigma}_{virt}$ in the $\overline{\text{MS}}$ scheme
  using the top quak mass evaluated at different choices of the renormalisation scale, 
  while the black solid line shows the on-shell result.}
\label{fig:renormalisationscheme}
\end{figure}

 We present our results in fig.~\ref{fig:renormalisationscheme}. We
 observe that indeed the $\Delta \hat{\sigma}_{virt}$ show a non-negligible
 dependence on the renormalisation scheme for the top quark mass. This
 holds true both for $gg\to HH$ and for $gg\to ZH$. In particular, for
 $gg\to ZH$ we see a shift of the maximal value of the differential
 $\Delta \hat{\sigma}_{virt}$, which is associated to the lower top
 quark mass value.  This may point to a significant uncertainty
 related to the top-mass scheme also in the case of $gg\to ZH$. We
 notice, however, that the effects observed in
 fig.~\ref{fig:renormalisationscheme} are likely to be partially
 compensated by similar modifications in the LO cross section and by
 the inclusion of the real-emission corrections at NLO. Therefore, we
 use fig.~\ref{fig:renormalisationscheme} only as an illustration of
 the flexibility of the merging method discussed in this paper, and we
 leave a full assessment of the effects due the change of the top-mass
 renormalization scheme to a future work.  For the same reason we
 refrain from providing a full comparison with ref.~\cite{Baglio:2020wgt}
 for $gg\to HH$.

\section{Conclusion}
\label{sec5}
In this paper, we  combined an expansion in small $\pt$ with a
HE expansion  for the processes $gg\to HH$ and $gg\to ZH$ and 
we showed that this combination leads to results that describe the whole phase space
of the considered processes with high accuracy.

The expansion in small $\pt$ is valid for $|\hat{t}| \lesssim 4
\mt^2$, while the HE expansion is valid for $|\hat{t}|\gsim 4 \mt ^2$
and $\hat{s}> 4 \mt^2$. For a successful combination of the two
expansion methods, we extended the validity range of both expansions
making use of simple Pad\'e approximants. This allowed us to describe
also the region $|\hat{t}| \simeq 4 \mt^2$ very accurately.

We  verified our method first at LO, where exact analytic results
are available. At the level of the differential partonic cross
section, the difference with the results in exact top quark mass
dependence never exceeded 1\% for any of the $M_{HH}$ bins.  In a
second step, we compared our results with a numerical grid containing
the results of a computation in exact top quark mass dependence for
the $gg\to HH$ process \cite{Davies:2019dfy}.  Also at NLO we could
verify the good numerical accuracy of our approach, observing
differences below 1\% for the invariant mass bins where the
numerical accuracy of the grid is assumed to be sufficiently high.

\newpage
We have also shown  that in our analytic calculation the top
quark mass renormalisation scheme can be easily changed. This shows
the great flexibility of our analytic approach with respect to the
numerical one. With a running time of well below 1 s per phase-space
point our results can be well implemented in a fast and versatile
Monte Carlo program.

\section*{Acknowledgements}
We thank Lina Alasfar and Xiaoran Zhao for useful discussions.
We would also like to thank the authors of ref.~\cite{Davies:2020drs}
 for a careful reading of the manuscript and for their useful comments.
G.D. would like to thank the Department of Physics of the University of Rome "La Sapienza" for the kind hospitality during part of this project.
The work of G.D. and M.V. was partially supported by the Italian Ministry of Research
(MUR) under grant PRIN 20172LNEEZ. The work of L.B. and 
P.P.G. has received financial support from Xunta de Galicia (Centro
singular de investigaci\'on de Galicia accreditation 2019-2022), by
European Union ERDF, and by ``Mar\'ia de Maeztu" Units of Excellence
program MDM-2016-0692 and the Spanish Research State Agency.

\bibliographystyle{utphys}
\bibliography{MergExp}

\end{document}